\documentclass[prd,twocolumn,showpacs,amsmath,amssymb]{revtex4}

\usepackage{amssymb}
\usepackage{mathrsfs}
\usepackage{txfonts}

\usepackage{graphicx}
\usepackage{dcolumn}% Align table columns on decimal point
\usepackage{bm}% bold math

\begin{document}

\title{Primordial non-Gaussianity in noncanonical warm inflation: three- and four-point correlations}

\author{Xiao-Min Zhang}
\email{zhangxm@mail.bnu.edu.cn}
\affiliation{School of Sciences, Qingdao University of Technology, Qingdao 266033, China}
\author{Hong-Yang Ma}
\email{hongyang_ma@aliyun.com}
\affiliation{School of Sciences, Qingdao University of Technology, Qingdao 266033, China}
\author{Peng-Cheng Chu}
\email{kyois@126.com}
\affiliation{School of Sciences, Qingdao University of Technology, Qingdao 266033, China}
\author{Jian-Yang Zhu}
\thanks{Corresponding author}
\email{zhujy@bnu.edu.cn}
\affiliation{Department of Physics, Beijing Normal University, Beijing 100875, China}

\date{\today}
%------------------------------------------------------------------------------------------------------------
\begin{abstract}
Non-Gaussianity generated in inflation can be contributed by two parts. The first part, denoted by $f_{NL}^{\delta N}$, is the contribution from four-point correlation of inflaton field which can be calculated using $\delta N$ formalism, and the second part, denoted by $f_{NL}^{int}$, is the contribution from the three-point correlation function of the inflaton field. We consider the two contributions to the non-Gaussianity in noncanonical warm inflation throughout (noncanonical warm inflation is a new inflationary model which is proposed in \cite{Zhang2014}). We find the two contributions are complementary to each other. The four-point correlation contribution to the non-Gaussianity is overwhelmed by the three-point one in strong noncanonical limit, while the conclusion is opposite in the canonical case. We also discuss the influence of the field redefinition, thermal dissipative effect and noncanonical effect to the non-Gaussianity in noncanonical warm inflation.

\end{abstract}
\maketitle

%------------------------------------------------------------------------------------------------------------
\section{\label{sec1}Introduction}
Inflation, as an necessary supplement to the standard model of the Universe, is an important branch of cosmology which can successfully solve the problems such as horizon, flatness and monopole \cite{Guth1981,Linde1982,Albrecht1982}. Another charming feature of inflation scenario is that it can give a natural mechanism to clarify the observed anisotropy of the cosmological microwave background (CMB) and the large scale structure exactly \cite{Weinberg,LiddleLyth,Dodelson}. Generally speaking, there are two kinds of inflationary theory till now: standard inflation, or sometimes called cold inflation, and warm inflation. Warm inflation was first proposed by A. Berera in 1995 \cite{BereraFang,Lisa2004,Berera2000}, and then has been developed a lot in the past twenty years, especially in the fields of perturbation theory \cite{Berera2000,BereraIanRamos,Lisa2004,MossXiong,Chris2009}, the micro-mechanism realization and dissipative issue of warm inflation \cite{MossXiong2006,Berera1999,BereraIanRamos}, and the consistency issue of warm inflation \cite{Ian2008,Campo2010,Zhang2015,Zhang2014,ZhangTachyon,ZhangZhu}. Standard and warm inflation share the advantages of solving horizon, flatness and monopole problems and generating nearly scale-invariant power spectrum. And warm inflation has its own advantages and improvements, such as curing the ``$\eta$" problem \cite{etaproblem} and the problem of overlarge amplitude of the inflaton suffered in some standard inflationary models \cite{Berera2005,BereraIanRamos}, and relaxing the strict slow roll conditions in standard inflation greatly. A most distinct difference between standard and warm inflation is the origin of density fluctuations. The cosmological perturbations can naturally arise from vacuum quantum fluctuations in standard inflation \cite{Weinberg,LiddleLyth,Dodelson,Bassett2006} while thermal fluctuations in warm inflation \cite{BereraFang,Lisa2004,Berera2000}. Warm inflation contains rich information about particle physics and can broad the scope of inflationary theory greatly. Some models that are already ruled out by new Planck observations \cite{PLANCKI2015} in standard inflation can again be in very good agreement with the Planck results in warm inflationary theory.

When studying inflation, one typically calculates the power spectrum of scalar perturbations and the amplitude of gravitational waves. These perturbation quantities, although very important, contain only two-point correlation statistics information. Two-point correlation information in perturbations is too limited to discriminate among a large range of inflationary models. There exists a so-called `degeneracy problem' (i.e. a single set of observables maps to a range of different inflation models) \cite{Eassona2013} in inflation theory. Even a precise measurement of the spectral index, the running of spectral index, and the detection of gravitational wave will not allow us efficiently discriminate among them. So we need the important information contained in primordial non-Gaussianity of inflation. The three-point function of curvature perturbation $\zeta$, or its Fourier transform, the bispectrum represents the lowest order statistics able to distinguish non-Gaussian from Gaussian perturbations \cite{Heavens1998,Ferreira1998}. In this paper we will concentrate on the lowest order non-Gaussianity. Non-Gaussianity contains useful message of inflation, which can help to distinguish different inflationary models.

Two-point correlation perturbations, i.e. power spectrums of scalar and tensor modes, generated in canonical standard inflation are already clear issues \cite{Weinberg,LiddleLyth,Dodelson,Bassett2006}. Many works also has been concentrated on the perturbations of noncanonical standard inflation. The research of scalar power spectrum, spectral index, the amplitude of gravitational wave and consistency relation shows that the sound speed, which is an character quantity describing noncanonical effect in noncanonical inflation, plays an important role in the two-point perturbation quantities \cite{Mukhanov19991,Mukhanov19992}. Non-Gaussianity, especially the three-point correlation in noncanonical standard inflation was researched in \cite{Creminelli2003,Tong2004,ChenHuang2007}, and these works found that a low sound speed can much enhance the level of non-Gaussianity. Many works calculate non-Gaussianity generated by multi-field inflation and reach the conclusion that multi-field inflation has more enhanced non-Gaussianity than single field inflation \cite{Vernizzi2006,Battefeld2007,Tower2010}. Non-Gaussianity in warm inflation was analysed specially from different opinion in some works \cite{MossXiong,Zhang2015,Zhang2016,MarGil2014,Gupta2002,Gupta2006}. In related works such as \cite{MossXiong,Gupta2006,Gupta2002,MarGil2014}, non-Gaussianity generated in canonical warm inflation was performed. Papers \cite{Gupta2002,Gupta2006} concentrated on the temperature independent warm inflationary case and \cite{MarGil2014,IanMoss2011} focused on the more complicated temperature dependent case. Thermal dissipation effect can increase non-Gaussianity to some extent. Canonical field was often used as inflaton in the research of warm inflation. Noncanonical warm inflation was first proposed in \cite{Zhang2014} and broaden the scope of inflationary picture. Non-Gaussianity in noncanonical warm inflation was first considered in our previous work \cite{Zhang2015}, and we get the result that small sound speed and large dissipation strength can both enhance the magnitude of non-Gaussianity. The works above all considered non-Gaussianity generated by inflaton fields in linear large-scale evolution of perturbations. More than ten years ago, $\delta N$ formalism, a gauge-invariant description of nonlinear curvature perturbation on large scales, was proposed to calculate the issue of non-Gaussianity \cite{Lyth2005,Zaballa2005,Vernizzi2006,Battefeld2007,Tower2010}. Nonlinear parameter $f_{NL}$ is often introduced to parameterize the magnitude of non-Gaussianity. Nonlinear parameter obtained by $\delta N$ formalism, i.e. $f_{NL}^{\delta N}$, is nearly scale independent, while nonlinear parameter generated by the intrinsic non-Gaussianities of inflaton fields in linear cosmological perturbation theory, i.e. $f_{NL}^{int}$ is often scale dependent. If the inflaton fields are Gaussian to sufficient accuracy, such as in canonical multi-field inflation, intrinsic result of non-Gaussianity $f_{NL}^{int}$ is overwhelmed by $\delta N$ result $f_{NL}^{\delta N}$ \cite{Sasaki2016,Vernizzi2006}. The two effects are complementary to each under field redefinition in standard inflation \cite{Sasaki2016}. Non-Gaussianity in canonical warm inflation was calculated from the $\delta N$ view in the work \cite{Zhang2016}, which is allowed by recent observations \cite{PLANCKNG2015}. That $f_{NL}^{\delta N}$ is less than one in large scale in canonical warm inflation is due to the overdamped thermal term, which can make the slow roll more easily to be satisfied.

In this paper we will analyse non-Gaussianity throughout in noncanonical warm inflation both from $\delta N$ view and intrinsic view. Since the intrinsic non-Gaussianity of inflaton field in noncanonical warm inflation is more prominent than in canonical inflation and the calculation of non-Gaussianity from $\delta N$ view is still absent, we'll calculate the $\delta N$ part non-Gaussianity, discuss the contributions to non-Gaussianity from both view and make comparisons between them. We also try to find how noncanonical effect and thermal effect influence the non-Gaussianity in noncanonical warm inflation. The paper is organized as follows: In Sec. \ref{sec2}, we introduce noncanonical warm inflationary scenario briefly and review the basic equations and important parameters of the new picture. In Sec. \ref{sec3}, we introduce non-Gaussian perturbation, $\delta N$ formalism and the evolution equations of inflaton perturbations in noncanonical warm inflation. Then we calculate the nonlinear parameter $f_{NL}$ from both $\delta N$ view and intrinsic view in noncanonical warm inflation concretely and give discussions of the non-Gaussian results respectively in Sec. \ref{sec4}. Finally, we draw the conclusions in Sec. \ref{sec5}.

\section{\label{sec2}The framework of noncanonical warm inflation}

Different from standard inflation, the scalar inflaton field is not isolated, but has interactions with other sub-dominated fields in warm inflation. Thanks to the interactions, a significant amount of radiation was produced constantly during the inflationary epoch, so the Universe is hot with a non-zero temperature $T$. There's a strong possibility that a warm Universe can happen \cite{Chris2008,Berera2016}.

The total matter action of the multi-component Universe in noncanonical warm inflation (noncanonical warm inflation is a kind of new inflationary model where noncanonical field behaves as inflaton \cite{Zhang2014}) is
\begin{equation}\label{action}
  S=\int d^4x \sqrt{-g} \left[ \mathcal{L}(\phi,X)+\mathcal{L}_R+\mathcal{L}_{int}\right],
\end{equation}
where $\mathcal{L}(\phi,X)$ is the Lagrangian density of the noncanonical inflaton field, $\mathcal{L}_R$ is the Lagrangian density of radiation fields and $\mathcal{L}_{int}$ denotes the interaction between the scalar fields. In the Friedmann-Robertson-Walker (FRW) Universe, the mean inflaton field is homogeneous, i.e. $\phi=\phi(t)$. Under some assumptions and calculations, we can get the evolution equation of the inflaton field by varying the action with respect to the inflaton field \cite{BereraFang,Berera1999,Zhang2014}:
\begin{equation}
 \mathcal{L}_{X}c_{s}^{-2}\ddot{\phi}+(3H\mathcal{L}_{X}+\Gamma)\dot{\phi}+V_{eff,\phi}(\phi,T)=0,  \label{EOMphi}
\end{equation}
where $H$ is the Hubble parameter which satisfies the Friedmann equation:
\begin{equation}\label{Friedmann}
  3H^2=8\pi G\rho.
\end{equation}
In Eq. (\ref{EOMphi}), $c_{s}^{2}=P_{X}/\rho_{X}=\left(1+2X\mathcal{L}_{XX}/\mathcal{L}_{X}\right)^{-1}$ is the sound speed which describes the traveling speed of scalar perturbations, $\Gamma$ is the dissipation coefficient and $V_{eff,\phi}(\phi,T)$ is the effective potential acquired thermal corrections. The subscripts $\phi$ and $X$ denote a derivative in our paper. The effective potential $V_{eff}(\phi,T)$ is different from the zero-temperature potential $V(\phi)$ in cold inflation. The thermal correction to the potential is constrained to be small enough by the slow roll conditions in warm inflation \cite{Ian2008,Campo2010,Zhang2014}. For simplicity we'll write $V_{eff}$ as $V$ hereinafter. The term $\Gamma \dot{\phi}$ in the evolution equation describes the dissipation effect of $\phi$ to radiations \cite{BereraFang,Berera2005,Berera2000,BereraIanRamos}, which is a thermal
damping term. In some papers, $\Gamma $ is often
set to be a constant for simplicity to analyse \cite{Herrera2010,Xiao2011,Taylor2000}.
Considering different microphysical basis of the interactions between inflaton and
other fields, different form of $\Gamma$ can been obtained \cite{MossXiong2006,MarGil2013,BereraIanRamos}. Generally speaking, $\Gamma$ can be a function of inflaton field and even Universe temperature.

An important parameter in warm inflationa is the dissipation strength which is defined as:
\begin{equation}\label{r}
  r=\frac{\Gamma}{3H}.
\end{equation}
This parameter describes the effectiveness of warm inflation, where $r\gg1$ refers to strong regime of warm inflation and $r\ll1$ refers to weak regime of warm inflation.

Thermal dissipative effect of warm inflation is accompanied by the production
of entropy. The expression for entropy density from thermodynamics is $s=-\partial f/\partial T$, and we have $s\simeq -V_T$ for that the free energy $f=\rho -Ts$ is dominated by potential during inflation.

The total energy density of the multi component Universe is
\begin{equation}
\rho =\frac 12\dot{\phi}^2+V(\phi ,T)+Ts.  \label{rho}
\end{equation}
and the total pressure is given by
\begin{equation}
p=\frac 12\dot{\phi}^2-V(\phi ,T).  \label{p}
\end{equation}
Combining the energy-momentum conservation
\begin{equation}
  \dot{\rho}+3H(\rho +p)=0, \label{conservation}
\end{equation}
with Eq. (\ref{EOMphi}), we can get the entropy production equation:
\begin{equation}
T\dot{s}+3HTs=\Gamma \dot{\phi}^2.  \label{entropy}
\end{equation}
The equation above is equivalent to the radiation energy density producing
equation $\dot{\rho}_r+4H\rho _r=\Gamma \dot{\phi}^2$, when the thermal correction to the effective potential is small enough in slow roll inflation.

Inflation is often associated with slow-roll approximation to drop the highest derivative terms in the equations of motion, thus we can get the slow roll equations of noncanonical warm inflation:
\begin{equation}
\dot{\phi}=-\frac{V_\phi}{3H(\mathcal{L}_{X}+r)},  \label{SRdotphi}
\end{equation}
\begin{equation}
Ts=r\dot{\phi}^2,  \label{SRTs}
\end{equation}
\begin{equation}
H^2=\frac{8\pi G}3 V,  \label{SRH}
\end{equation}
\begin{equation}
4H\rho _r=\Gamma \dot{\phi}^2.  \label{SRrho}
\end{equation}
The validity of the slow roll approximation depends on the slow roll conditions given by systemic stability analysis \cite{Ian2008,Campo2010,Lisa2004,Zhang2014}. The slow roll conditions are associated with some important slow roll parameters defined as
\begin{equation}
\epsilon =\frac{M_p^2}{2}\left(\frac{V_{\phi}}{V}\right) ^2, \eta =M_p^2\frac {V_{\phi \phi}}{V}, \beta
=M_p^2\frac{V_{\phi}\Gamma_{\phi}}{V\Gamma},
\end{equation}
When dealing with warm inflation, we'll need two additional slow roll parameters:
\begin{equation}
b=\frac {TV_{\phi T}}{V_{\phi}}
\end{equation}
and
\begin{equation}
c=\frac{T\Gamma_T}{\Gamma}
\end{equation}
to describe the temperature dependence of effective potential and dissipation coefficient in warm inflation \cite{Ian2008,Campo2010}.
These slow roll parameters are potential slow roll (PSR) parameters, which have relations with inflation potential and are different from Hubble slow roll (HSR) parameters. HSR parameters are invariant under field redefinition while PSR parameters are not.

The slow-roll approximations can be guaranteed when
\begin{eqnarray}
\epsilon\ll\frac{\mathcal{L}_{X}+r}{c^2_s},\beta\ll\frac{\mathcal{L}_{X}+r}{c^2_s},\eta\ll\frac{\mathcal{L}_{X}}{c^2_s},
\nonumber \\ b\ll\frac{min\{\mathcal{L}_{X},r\}}{(\mathcal{L}_{X}+r)c^2_s},~~~~|c|<4~~~ , \label{SRcondition}
\end{eqnarray}
in noncanonical warm inflationary scenario \cite{Zhang2014}. The additional parameter $c$ is not necessarily small, but a stability analysis of warm inflation shows that $|c|<4$ for a consistent model \cite{Ian2008,ZhangZhu,Zhang2014}. These slow roll conditions are more easy to be satisfied than in canonical warm inflation, let alone standard inflation.
 The number of e-folds in warm inflation is given by
\begin{equation}
N=\int H dt=\int\frac{H}{\dot{\phi}}d\phi\simeq-\frac{1}{M_p^2}\int_{\phi_{\ast}}
^{\phi_{end}}\frac{V(\mathcal{L}_X+r)}{V_{\phi}}d\phi,  \label{efold}
\end{equation}
where $M_p^2=\frac 1{8\pi G}$.

\section{\label{sec3}Non-Gaussian perturbations of inflation}

\subsection{\label{sec31}$\delta N$ formalism and non-Gaussianity}

Now we'll give a brief introduction of $\delta N$ formalism and primordial non-Gaussianity of inflation. $\delta N$ formalism was proposed in \cite{Lyth2005,David2005,Starobinsky,Sasaki1996,Sasaki1998} and then often used in calculating the non-Gaussianity of double and multi-field inflationary models \cite{Vernizzi2006,Battefeld2007,Tower2010}.

The primordial curvature perturbation on uniform density hypersurfaces of the Universe, denoted by $\zeta$, is already present a few Hubble times before cosmological scales start to enter the horizon. And observations suggest the perturbation $\zeta$ was Gaussian term dominated with a nearly scale-invariant spectrum.

Considering small perturbations in the background of the flat FRW Universe with scale factor $a(t)$, the spatial metric is given by
\begin{equation}\label{gij}
  g_{ij}=a^2(t)e^{2\zeta(t,\mathbf{x})}\gamma_{ij}(t,\mathbf{x})=\tilde{a}^2(t,\mathbf{x})\gamma_{ij}(t,\mathbf{x}),
\end{equation}
where $\gamma_{ij}(t,\mathbf{x})$ has unit determinant and accounts for the tensor perturbation. We can find that according to this definition, $\zeta$ is the perturbation in $\ln \tilde{a}$.

According to $\delta N$ formalism \cite{Lyth2005,David2005,LythMalik,Starobinsky,Sasaki1996,Sasaki1998}, $\zeta$, evaluated at some time $t$, is equivalent to the perturbation of the number of e-foldings $N(t,\mathbf{x})$ from an initial flat hypersurface at $t=t_{in}$, to a finial uniform density or, equivalently, comoving hypersurface at the time of $t$. Thus we have
\begin{equation}\label{zeta}
  \zeta(t,\mathbf{x})=\delta N \equiv N(t,\mathbf{x})-N_0(t),
\end{equation}
where $N(t,\mathbf{x})\equiv\ln[\frac{\tilde{a}(t)}{a(t_{in})}]$ and $N_0(t)\equiv\ln[\frac{a(t)}{a(t_{in})}]$.

The evolution of the Universe is supposed to be determined mainly by one or more inflaton fields during inflationary epoch. Choosing the convenient flat slicing gauge and considering perturbations, we can expand each scalar field in the form $\Phi_i(t,\mathbf{x})=\phi_i(t)+\delta\phi_i(t,\mathbf{x})$. As mentioned above, the curvature perturbation $\zeta$ is almost Gaussian, so we can expand $\zeta$ up to second order for good accuracy:
\begin{equation}\label{zeta2}
  \zeta(t,\mathbf{x})=\delta N\simeq \sum_i N_{,i}(t)\delta\phi_i+\frac12\sum_{ij} N_{,ij}(t)\delta\phi_i\delta\phi_j,
\end{equation}
where $N_{,i}\equiv\frac{\partial N}{\partial\phi_i}$ and $N_{,ij}\equiv\frac{\partial^2 N}{\partial\phi_i\partial\phi_j}$. They may be entirely responsible for any observed non-Gaussianity if the field perturbations are pure Gaussian, which are the contributions of four-point correlations. However, the inflaton field perturbation in noncanonical warm inflation deviates from pure Gaussian distribution to some extent that larger than in canonical inflation. Thus we also need to compute non-Gaussianity generated by intrinsic non-Gaussianity of inflaton field, i.e. the three-point correlations of field.

The power spectrum of the curvature perturbation $\zeta$, denoted by $\mathcal{P}_{\zeta}$, is defined as
\begin{equation}\label{spectrum}
  \langle\zeta_{\mathbf{k}_1}\zeta_{\mathbf{k}_2}\rangle\equiv(2\pi)^3\delta^3(\mathbf{k}_1+\mathbf{k}_2)
\frac{2\pi^2}{k_1^3}\mathcal{P}_{\zeta}(k_1),
\end{equation}
and $\mathcal{P}_{\zeta}(k)\equiv\frac{k^3}{2\pi^2}P_{\zeta}(k)$.

The lowest order non-Gaussianity is three-point function of curvature perturbation, or its Fourier transform, the bispectrum, which is defined through
\begin{equation}\label{bispectrum}
  \langle\zeta_{\mathbf{k}_1}\zeta_{\mathbf{k}_2}\zeta_{\mathbf{k}_3}\rangle\equiv(2\pi)^3\delta^3
(\mathbf{k}_1+\mathbf{k}_2+\mathbf{k}_3)B_{\zeta}(k_1,k_2,k_3).
\end{equation}
Its normalization is specified by the nonlinear parameter $f_{NL}$ through
\begin{equation}\label{fnl}
  B_{\zeta}(k_1,k_2,k_3)\equiv -\frac65f_{NL}(k_1,k_2,k_3)\left[P_{\zeta}(k_1)P_{\zeta}(k_2)+cyclic \right].
\end{equation}
Observational limits are usually put on the nonlinear parameter and it is often used to describe the level of non-Gaussianity effectively. We can compute the bispectrum through Eq. (\ref{zeta2}) and the calculation can yield
\begin{eqnarray}\label{threepoint}
  \langle\zeta_{\mathbf{k}_1}\zeta_{\mathbf{k}_2}\zeta_{\mathbf{k}_3}\rangle =\sum_{ijk}N_{,i}N_{,j}N_{,k}
  \langle\delta\phi^{i}_{\mathbf{k}1}\delta\phi^{j}_{\mathbf{k}2}\delta\phi^{k}_{\mathbf{k}3}\rangle ~~~~~~~~~~~\nonumber\\
  +\frac12\sum_{ijkl}N_{,i}N_{,j}N_{,kl}\langle\delta\phi^{i}_{\mathbf{k}1}\delta\phi^{j}_{\mathbf{k}2}
  (\delta\phi^{k}\star\delta\phi^{l})_{\mathbf{k}3}\rangle+perms,
\end{eqnarray}
where a star denotes the convolution and the correlation functions higher than four-point are neglected. The first line in the equation above, a three-point correlation, is the contribution from the intrinsic non-Gaussianity of the inflaton fields, which can be scale dependent; while the second line, a four-point correlation, is scale independent and can be calculated conveniently by using $\delta N$ formalism. Based on the $\delta N$ formalism, we can get the part of non-Gaussianity generated by four-point correlation. The expression of $\delta N$ part nonlinear parameter is given by \cite{Lyth2005,Boubekeur}:
\begin{equation}\label{fNL}
  -\frac35 f_{NL}^{\delta N}=\frac{\sum_{ij}N_{,i}N_{,j}N_{,ij}}{2\left[\sum_i N^2_{,i}\right]^2}.
\end{equation}
We can see that the $f_{NL}^{\delta N}$ term is scale independent. The total non-Gaussianity should be described by $f_{NL}=f_{NL}^{\delta N}+f^{int}_{NL}$.

\subsection{\label{sec32}thermal fluctuations of inflaton field}

In noncanonical warm inflation, there is only one scalar field acting as inflaton and we can expand the full inflaton as $\Phi(\mathbf{x},t)=\phi(t)+\delta\phi(\mathbf{x},t)$, where $\delta\phi$ is the small perturbation around the homogenous background field $\phi(t)$ as usual. The evolution of inflaton is in overdamped regime in noncanonical warm inflation due to the enhanced Hubble damping term and thermal damping term. The evolution of the inflaton perturbations is very slow in slow roll regime as indicated in \cite{Lisa2004}, so the evolution equation of full inflaton in slow roll noncanonical warm inflation can be given by \cite{Berera2000,Taylor2000,Zhang2014}:
\begin{equation}\label{EOMphik}
\frac{d\Phi(\mathbf{k},t)}{dt}=\frac{1}{3H\mathcal{L}_X+\Gamma}\left[-k^2\mathcal{L}_X\delta\phi(\mathbf{k},t)-V_{\phi}
(\Phi(\mathbf{k},t))+\xi(\mathbf{k},t)\right],
\end{equation}
where $\xi$ is the thermal stochastic noise in thermal system with zero mean $\langle\xi\rangle=0$. In
the high temperature limit $T\rightarrow\infty$, the noise source is Markovian: $\langle\xi(\mathbf{k},t)\xi(\mathbf{k'},t')\rangle=2\Gamma T(2\pi)^3\delta^3(\mathbf{k}-\mathbf{k'})\delta(t-t')$ \cite{Lisa2004,Gleiser1994}. Thermal noise term in warm inflation is a kind of Gaussian distributed white noise \cite{Berera2000}. Since the leading order inflaton perturbation is linear response to the thermal noise, it is also Gaussian distributed. So if we want to calculate the predicted bispectrum of inflaton perturbation from Eq. (\ref{EOMphik}), we should expand the inflaton fluctuations to second order at least: $\delta\phi(\mathbf{x},t)=\delta\phi_1(\mathbf{x},t)+\delta\phi_2(\mathbf{x},t)$, where $\delta\phi_1=\mathcal{O}(\delta\phi)$ and $\delta\phi_2=\mathcal{O}(\delta\phi^2)$. Then the equations of motion for the first and second order fluctuations in Fourier space can be obtained from Eq. (\ref{EOMphik}):
\begin{eqnarray}\label{deltaphi1}
\frac{d}{dt}\delta\phi_1(\mathbf{k},t)&=&\frac{1}{3H\mathcal{L}_{X}+\Gamma}\left[-\mathcal{L}_Xk^2\delta\phi_1(\mathbf{k},t)\right.\nonumber\\
&-&\left.V_{\phi\phi}(\phi(t))\delta\phi_1(\mathbf{k},t)+\xi(\mathbf{k},t)\right],
\end{eqnarray}
\begin{eqnarray}\label{deltaphi2}
\frac{d}{dt}\delta\phi_2(\mathbf{k},t)&=&\frac{1}{3H\mathcal{L}_{X}+\Gamma}\left[-\mathcal{L}_Xk^2\delta\phi_2(\mathbf{k},t)-V_{\phi\phi}
(\phi(t))\delta\phi_2(\mathbf{k},t)\right.\nonumber\\ &-&\left.\frac12V_{\phi\phi\phi}(\phi(t))\int\frac{dp^3}{(2\pi)^3}\delta\phi_1(\mathbf{p},t)\delta\phi_1(\mathbf{k}
-\mathbf{p},t)\right. \nonumber\\ &-& \left. k^2\mathcal{L}_{XX}\int\frac{dp^3}{(2\pi)^3}\delta\phi_1(\mathbf{p},t)\delta X_1(\mathbf{k}
-\mathbf{p},t)\right],
\end{eqnarray}

The equation of motion for the fluctuations is obtained through perturbing the evolution equation of the full inflaton to second order. The analytic solutions of first and second order fluctuations, $\delta\phi_1$ and $\delta\phi_2$, can be obtained by solving the two equations above. And then the non-Gaussianity generated by intrinsic non-Gaussian distributions of inflaton can be obtained \cite{Zhang2015}, which we'll analyse concretely in next section.

\section{\label{sec4}Non-Gaussianity in noncanonical warm inflation}

There's only one inflaton field in noncanonical warm inflation, so only one $\delta\phi_i$ is relevant, then Eq. (\ref{zeta2}) reduces to
\begin{equation}\label{zeta3}
  \zeta(t,\mathbf{x})=N_{,i}\delta\phi_i+\frac12 N_{,ii}\left(\delta\phi_i\right)^2,
\end{equation}
so we can get
\begin{equation}\label{fNL1}
  -\frac35f_{NL}^{\delta N}=\frac12\frac{N_{,ii}}{N^2_{,i}}.
\end{equation}
Since there is only one $\delta\phi_i$, without ambiguity, we can rewrite $N_{,i}$ as $N_{\phi}$ and $N_{,ii}$ as $N_{\phi\phi}$ below.

Through Eq. (\ref{efold}), we can get
\begin{equation}\label{Nphi}
  N_{\phi}=-\frac{1}{M_p^2}\frac{V(\mathcal{L}_{X}+r)}{V_{\phi}},
\end{equation}

Observational limits of primordial non-Gaussianity generated by inflation are usually put on the nonlinear parameter. And it's estimated on the time of horizon crossing, which is well inside the slow roll inflationary regime. So we'll first calculate the $\delta N$ part nonlinear parameter $f_{NL}^{\delta N}$ in slow roll approximation.
Now we will consider a general dissipative coefficient case with $\Gamma=\Gamma(\phi,T)$ in slow roll noncanonical warm inflation.

From Eq. (\ref{Nphi}), we can get
\begin{equation}\label{Nphiphi1}
  N_{\phi\phi}=-\frac{1}{M_p^2}\left[(\mathcal{L}_{X}+r)-\frac{(\mathcal{L}_{X}+r)\eta}{2\epsilon}-\frac r2+\frac{\beta r}{2\epsilon}\right].
\end{equation}
Then we can obtain the $\delta N$ part nonlinear parameter in noncanonical warm inflation with a general dissipation coefficient from Eqs (\ref{fNL1}), (\ref{Nphi}) and (\ref{Nphiphi1}):
\begin{equation}\label{fNLdeltaN}
  f_{NL}^{\delta N}=\frac{5\epsilon}{3(\mathcal{L}_{X}+r)}-\frac{5\eta}{6(\mathcal{L}_{X}+r)}-\frac{5r\epsilon}
  {6(\mathcal{L}_{X}+r)^2}+\frac{5r\beta}{6(\mathcal{L}_{X}+r)^2}.
\end{equation}

As $\delta N$ formalism indicated, the $\delta N$ part nonlinear parameter $f_{NL}^{\delta N}$ is scale independent, for it can be decided only by nonperturbative background equations. Considering the slow roll conditions Eq. (\ref{SRcondition}) in noncanonical warm inflation, we can find from the equation above that $|f_{NL}^{\delta N}|\sim \mathcal{O}\left(\epsilon/(\mathcal{L}_{X}+r)\right)\lesssim1$, which is a first order small quantity in slow roll approximation. As the slow roll conditions suggest, during the inflationary epoch, the amplitude of non-Gaussianity is quite small and can grow slightly along with the inflation of Universe. Thus the level of non-Gaussiaity generated by four-point correlation in noncanonical warm inflation, characterized by parameter $f_{NL}^{\delta N}$, is not significant as in canonical warm inflation. Since the $\delta N$ form non-Gaussianity is not large enough, it's unsafe to use this part to represent the whole primordial non-Gaussianity generated by inflation as some papers performed \cite{Zhang2016,David2005}. So the calculation of non-Gaussianity generated by three-point correlation functions of inflaton field is also necessary.

Non-Gaussianity generated by intrinsic non-Gaussianity of inflaton, characterized by $f_{NL}^{int}$, is considered in some papers \cite{Gupta2002,Gupta2006,MossXiong,Zhang2015,Zhang2016,MarGil2014}. Paper \cite{Zhang2015} researched non-Gaussianity in noncanonical warm inflation with a temperature independent dissipative coefficient preliminarily and yields:
\begin{equation}\label{fNLint}
f_{NL}^{int}=-\frac56
\ln\sqrt{\frac{3(\mathcal{L}_X+r)}{\mathcal{L}_X}}\left[\frac{\epsilon\varepsilon}{(\mathcal{L}_X+r)^2}
+\left(\frac{1}{c_s^2}-1\right)\right].
\end{equation}
The intrinsic part nonlinear parameter $f_{NL}^{int}$ is estimated for the three wavenumber $\mathbf{k}_1$, $\mathbf{k}_2$, $\mathbf{k}_3$ all within a few e-folds of exiting the horizon, i.e. there is a mild hierarchy among them. The intrinsic nonlinear parameter $f_{NL}^{int}$ is weakly dependent on the time and different wavenumbers, so it has a good scale independent approximation. We can see that $f_{NL}^{int}$ can be much greater than one, and thereby much greater than $f_{NL}^{\delta N}$ when the sound speed $c_s$ is low. A low sound speed can much enhance the magnitude of primordial non-Gaussianity in noncanonical warm inflation, thus $f_{NL}$ should be dominated by $f_{NL}^{int}$ term. The first term in Eq. (\ref{fNLint}) is a second order small quantity, while $|f_{NL}^{\delta N}|$ is a first order small quantity, so the first term in $f_{NL}^{int}$ is absolutely overwhelmed by the second term and can be neglected. Then $f_{NL}^{int}\cong -\frac56\left(\frac{1}{c_s^2}-1\right)\ln\sqrt{\frac{3(\mathcal{L}_X+r)}{\mathcal{L}_X}}~\sim
\mathcal{O}\left(\frac{1}{c_s^2}-1\right)$, which suggests both strong noncanonical effect (characterized by low sound speed $c_s$) and strong dissipative strength (characterized by large $r$) contribute to large magnitude of non-Gaussianity, but obviously the contribution of low sound speed is more significant.

The whole non-Gaussianity should be described by the nonlinear parameter $f_{NL}=f_{NL}^{\delta N}+f_{NL}^{int}$. The two part are complementary to each other and both are not invariant under field redefinition. We can find that the non-Gaussianity in noncanonical warm inflation is dominated by intrinsic non-Gaussianity of inflaton field from the discussions above. The term $f_{NL}^{int}$ cannot be overlooked, instead it plays important role in non-Gaussian problems of noncanonical warm inflation. We know that PSR parameters are all related to inflaton field and thus they are not invariant quantities under field redefinition, while Hubble parameter $H$, HSR parameters, sound speed $c_s$ are invariant quantities. When a Lagrangian density $\mathcal{L}(X,\phi)$ is given, the parameters $X$ and thus $\mathcal{L}_{X}$ are both variant under field redefinition. The important characteristic parameters in warm inflation, $\Gamma$, and thus $r$ are also variant under field redefinition. Through Eqs. (\ref{fNLdeltaN}) and (\ref{fNLint}), we can see that both $f_{NL}^{\delta N}$ and $f_{NL}^{int}$ are not invariant under field redefinition. While the total $f_{NL}$ should be an invariant quantity under field redefinition, so the two parts $f_{NL}^{\delta N}$ and $f_{NL}^{int}$ are complementary to each other. Since the calculation of $f_{NL}^{int}$ part is more complicated especially in noncanonical warm inflation, we can try to choose an appropriate field gauge to simplify the calculation of total non-Gaussianity to some extent. The non-Gaussian result can be well inside the region allowed by Planck observations \cite{PLANCKNG2015} when $c_s$ is not small enough.

The non-Gaussian results in noncanonical warm inflation can reduce to canonical case when $c_s\rightarrow1$:
\begin{equation}\label{fnlncanonical}
  f_{NL}^{\delta N}=\frac{5\epsilon}{3(1+r)}-\frac{5\eta}{6(1+r)}-\frac{5r\epsilon}{6(1+r)^2}+\frac{5r\beta}{6(1+r)^2},
\end{equation}
and
\begin{equation}\label{fnlintcanonical}
  f_{NL}^{int}=-\frac56\ln\sqrt{3(1+r)}\frac{\epsilon\varepsilon}{(1+r)^2},
\end{equation}
where $\varepsilon=2M_p^2\frac{V_{\phi\phi\phi}}{V_{\phi}}$ can be seen as a first order slow-roll small quantity which has the same magnitude as the slow-roll parameter $\eta$ in the monomial potential case. In canonical warm inflation, the situation is quite different from that in noncanonical warm inflation. In canonical warm inflation, $f_{NL}=f_{NL}^{\delta N}+f^{int}_{NL}$ is dominated by the $f_{NL}^{\delta N}$ term, since the term $f^{\delta N}_{NL}$ is a first order slow roll small quantity while $f_{NL}^{int}$ is a second order small quantity. From Eqs. (\ref{fnlncanonical}) and (\ref{fnlintcanonical}), we can see that primordial non-Gaussianity in canonical warm inflation is not significant, which is quite different from noncanonical case. The intrinsic non-Gaussian results represented by Eqs. (\ref{fNLint}) and (\ref{fnlintcanonical}) are obtained in the warm inflationary case with a temperature independent dissipative coefficient $\Gamma=\Gamma(\phi)$. The non-Gaussianity generated in canonical warm inflation with a more complicated temperature dependent dissipative coefficient $\Gamma=\Gamma(\phi,T)$ is considered in \cite{IanMoss2011}, where the authors found that the non-Gaussianity can be significant to some extent. In temperature dependent case, the inflaton and radiation fluctuations are coupled to each other, the analytic result for power spectrum is too hard to be obtained, and we can only get a numerical result \cite{Chris2009}. The coupling between inflaton fluctuations and radiation fluctuations can make warm inflation stronger in most cases and also enhance the magnitude of non-Gaussianity \cite{IanMoss2011}. Some papers proposed a general form of the dissipative coefficient in warm inflation like $\Gamma=C_{\phi}\frac{T^m}{\phi^{m-1}}$ \cite{Zhangyi2009} and $\Gamma=\Gamma_0(\frac{\phi}{\phi_0})^n(\frac{T}{\tau_0})^m$ \cite{Lisa2004,Campo2010}, where $n$ and $m$ are integers. In these general forms, the characteristic warm inflationary parameter $c=m$, and the warm inflationary case can reduce to temperature independent case when $c=0$. The non-Gaussianity in canonical temperature dependent warm inflation can depend on the parameter $c$ by a function $f(c)$ \cite{IanMoss2011}, where the function $f(c)$ is greater than 1 in most cases (i.e. the cases with $c>0$) while reduce to 1 when $c=0$. We can see that the coupling between inflaton fluctuations and radiation fluctuations can enhance the magnitude of non-Gaussianity in canonical warm inflation, the result should still hold qualitatively (i.e. the magnitude of non-Gaussianity in temperature dependent noncanonical warm inflation is enhanced by a factor $f(c)>1$ compared to temperature independent case) in noncanonical warm inflation with a different and more complicated form of $f(c)$. The quantitative analysis is complicated to some extent and we'll concentrate completely on this problem in our next work.

\section{\label{sec5}conclusions and discussions}

In this paper, we investigate the whole primordial non-Gaussianity generated in noncanonical warm inflation. We give a brief introduction of noncanonical warm inflationary theory. Non-Gaussianity generated by inflation is often described by nonlinear parameter $f_{NL}$ and it can be divided into two parts: $f_{NL}^{\delta N}$ and $f_{NL}^{int}$. The first part describes the contribution of the four-point correlation of inflaton perturbation and the second part is due to the three-point correlation, i.e. the intrinsic non-Gaussianity of inflaton field. The two parts are complementary to each and they together can describe the primordial non-Gaussianity in inflation entirely. In addition, the two parts are both variant under field redefinition, while the whole $f_{NL}$ should be invariant.

$\delta N$ formalism is convenient to use and so is often used in calculating the non-Gaussianities in multi-field inflation theories. We introduce $\delta N$ formalism and the evolution equation of the perturbation of inflaton field in noncanonical warm inflation briefly. Noncanonical warm inflation is dominated by one inflaton field, so we use the $\delta N$ formalism that reduces to single field case to calculate the parameter $f_{NL}^{\delta N}$. The $\delta N$ part nonlinear parameter $f_{NL}^{\delta N}$ in noncanonical warm inflation is scale-independent. Using $\delta N$ formalism, we obtain the expression of $f_{NL}^{\delta N}$ in noncanonical warm inflation with a general coefficient $\Gamma=\Gamma(\phi,T)$. We reach the conclusion that $f_{NL}^{\delta N}$ can be expressed as a linear combination of the PSR parameters, so it's a first order small quantity in
slow roll approximation with the order $|f_{NL}^{\delta N}|\sim \mathcal{O}\left(\frac{\epsilon}{\mathcal{L}_{X}+r}\right)$. That indicates the $\delta N$ part non-Gaussianity generated by noncanonical warm inflation is insignificant as in canonical single-field inflation. Since the magnitude of non-Gaussianity represented by $f_{NL}^{\delta N}$ is not large enough, it's unsafe to ignore the non-Gaussianity comes from the self-interaction of the inflaton field. So we also consider the non-Gaussianity generated by intrinsic non-Gaussianity of inflaton, which is represented by $f_{NL}^{int}$, and find that this part overwhelm the $f_{NL}^{\delta N}$ part. In noncanonical warm inflation, sub-light sound speed of noncanonical inflaton contribute mostly to the non-Gaussianity, and thermal dissipation effect and high order correlations also contribute to non-Gaussianity to a certain extent.

%----------------------------------------------------------------------------------------------------
\acknowledgments This work was supported by the National Natural Science Foundation of China (Grants No. 11605100, 11547035, 11505100, 11235003 and 11575270).
%---------------------------------------------------------------------------------------------------------

\end{document}